\documentclass[a4paper]{jpconf}
\usepackage{amssymb,graphicx}

\newcommand{\R}{\mathbb{R}}
\newcommand{\C}{\mathbb{C}}
\newcommand{\hil}{\mathcal{H}}

\newcommand{\bsigma}{\mbox{\boldmath $\sigma$}}

\newcommand{\bham}{\mbox{\boldmath $H$}}
\newcommand{\buni}{\mbox{\boldmath $U$}}

\newcommand{\bldm}{\mbox{\boldmath $M$}}
\newcommand{\state}[1]{\vert #1\rangle}

\newtheorem{defn}{Definition}
\newtheorem{example}{Example}

\begin{document}

\title{Classicality in Quantum Mechanics}

\author{Olaf Dreyer}
\address{Theoretical Physics, Blackett Laboratory\\
Imperial College London, London, SW7 2AZ, U.K.}
\ead{o.dreyer@imperial.ac.uk}

\begin{abstract} In this article we propose a solution to the measurement problem in quantum mechanics. We point out that the measurement problem can be traced to an a priori notion of classicality in the formulation of quantum mechanics. If this notion of classicality is dropped and instead classicality is defined in purely quantum mechanical terms the measurement problem can be avoided. We give such a definition of classicality. It identifies classicality as a property of large quantum system. We show how the probabilistic nature of quantum mechanics is a result of this notion of classicality. We also comment on what the implications of this view are for the search of a quantum theory of gravity.
\end{abstract}

\section{Introduction}
One of the striking features of ordinary quantum mechanics is its linear structure. Given two possible states of a system there is a notion of the sum of these two states. No such notion exists in classical mechanics. The phase space of classical systems does not have the structure of a vector space. In general we can not add the states of a classical system. Since quantum mechanics is the more fundamental theory the question arises of how this non-linear property of classical physics arises from the linear theory. This problem is usually called the measurement problem since it is in the process of a measurement on the system that the problem requires a solution. The solution to the problem is usually referred to as an interpretation. In this article we argue that no additional interpretation of ordinary quantum mechanics is required if only one carefully defines what is meant by a classical state. 

The relation of classicality to quantum mechanics is often blurred by the fact that we have constructed the quantum theory from the classical theory by a process of quantization. This means that the states of the classical system are also possible states of the quantum system. We regard this fact as one of the major reason for why the measurement problem exists. If the classical states are known from the outset the question arises of how the quantum system assumes one of these states. Since the quantum system might be in a superposition of these classical states the measurement problem can not be escaped.

In our presentation we thus want to be more careful. We present quantum mechanics and its relation to classical mechanics in three steps. In the first step (section \ref{sec:qm}) we present the basic setup of quantum mechanics. Here we adhere to the usual linear structure. The state space is given by a Hilbert space and the evolution is given by a unitary one-parameter group generated by a self-adjoint Hamiltonian. We will not introduce any notion of classicality at this point and we do not talk about measurements yet. In the second step (section \ref{sec:class}) we then introduce the notion of a classical property. The important point here is that it is a quantum system that has this classical property. No notion of classicality outside of quantum mechanics needs to be introduced. In the last step (section \ref{sec:trans}) we then look at the transition from quantum mechanical states to classical states. This is where one usually encounters the measurement problem. We will show how with our definition of classicality  the measurement problem does not arise. In the concluding section \ref{sec:concl} we comment on what these views of quantum mechanics mean for the search for a quantum theory of gravity. 

\section{First step: Quantum mechanics}\label{sec:qm}
The basic setup we will use is that of standard quantum mechanics. We thus start with a Hilbert space $\hil$. An example of such a Hilbert space could be the space of $N$ qbits: 
\begin{equation}
\hil = (\C^2)^{\otimes N}
\end{equation}
On this Hilbert space the evolution of the system is described by a Hamiltonian $\bham$. It generates time evolution through the unitary 
\begin{equation}
\buni = \exp{i\bham t}.
\end{equation}
This is the basic setup we will use. It is completely linear and deterministic. We will not add a collapse postulate that breaks the linearity. We will also not give a probabilistic interpretation of the states in $\hil$ or postulate the Born rule.

The most important difference to the standard formulation of quantum mechanics is that we do not have an a priori notion of classicality. When setting up a quantum mechanical calculation the Hilbert space is often of the form
\begin{equation}
L^2( \mbox{classical states} ).
\end{equation}
An example is $L^2(\R)$, where $\R$ represents the space of positions of a particle. We do not want to start with such an identification of what a classical state is. The reason is that such an identification forces the measurement problem on us. If one assumes it from the outset the question for how the classical world arises is fixed: How does the state of the system become
\begin{equation}
\state{\mbox{classical state}}\mbox{?}
\end{equation}
Given that the initial state of the system was a superposition of states of this form one encounters the measurement problem. 

The inclusion of a classical world into the basic formulation of the theory is also unsatisfactory in that a less fundamental theory is needed to formulate the more fundamental one. We want to avoid this mix of ontologies and start with a quantum theory that has no notion of classicality built in.

\section{Second step: Classical states}\label{sec:class}
Given that we have no a priori notion of classicality we now have to define what we mean by classicality. The definition we will use is the following:

\begin{defn} A system has a \emph{classical property} $\theta_0$ if the free energy describing the system is of the form
\begin{equation}
F(\theta_0 + \Delta\theta) = F(\theta_0) + c(\Delta\theta)^2,
\end{equation}
for some constant $c>0$. 
\end{defn} 

If a system has a classical property $\theta_0$ we can introduce a generalized force 
\begin{equation}
f = \frac{\delta F}{\delta\Delta \theta}.
\end{equation}
In solid state physics this force is referred to as generalized elasticity or rigidity. 

\begin{example} A large class of examples of systems with classical properties is provided by systems that have undergone symmetry breaking. The order parameter that is non-zero in the symmetry broken phase is the classical property. The easiest example here is spin chain where the classical property is the magnetization $\bldm$. 
\end{example}

\begin{example} Symmetry breaking does not exhaust the class of examples of classical properties. The formation of a surface in the gas to liquid transition is an example that does not fit neatly into the symmetry breaking paradigm. Here the classical property is the surface $\bsigma$. The corresponding generalized force is the surface tension. 
\end{example}

Defined this way we see that a classical property is a \emph{property of a large quantum system}. 

The presence of the generalized force $f$ makes the classical property of the system observation independent. The interaction with a system that has a classical property $\theta$ is effectively described using this classical property. To infer what the magnetization $\bldm_1$ of a spin chain is one might probe it with another spin chain with magnetization $\bldm_2$. Their interaction is best described by a term of the form $\bldm_1\cdot\bldm_2$. This probing does not destroy the classical property. It becomes in this sense an objective, observer and observation independent, property. 

\section{Third step: Transition}\label{sec:trans}
With this new definition of classicality we can now ask the question what happens when a classical property emerges. The key observation is that this transition from quantum to classical is discontinuous and very sensitive to the environment. We claim that it is the statistic nature of the environment that enters here that is responsible for the probabilistic nature of quantum mechanics. 

\begin{figure}
\begin{center}
\includegraphics{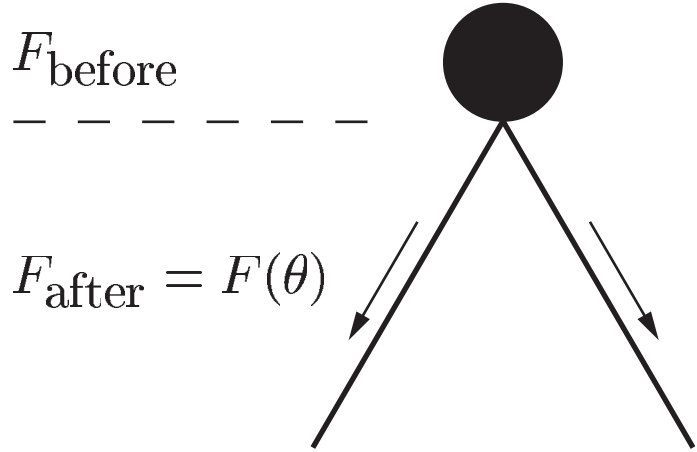}
\caption{A crude classical picture illustrating the nature of the quantum to classical transition. Before the transition the free energy $F$ of the system does not depend on $\theta$. In the transition it acquires a dependence on $\theta$. This transition is naturally discontinuous and very sensitive to the state of the environment as is the motion of the ball in this classical image. It is here that the probabilistic nature of quantum mechanics emerges.}\label{fig:balance}
\end{center}
\end{figure}

Before the system acquires a classical property $\theta_0$ the free energy describing the system does not depend on $\theta$. After the system has acquired the classical property $\theta_0$ it depends on $\theta$.This change in behavior of the free energy is by necessity discontinuous. It is also very sensitive to the state of the environment. 

We see that the environment has three roles that are not usually acknowledged. Because the system after the transition is in a state of lower entropy and energy the environment functions as a dump for the excess energy and entropy. The environment is further required to bring the system close to a transition point. This might be achieved by adjusting pressure or temperature to appropriate values. Finally the environment is responsible for the stochastic character of the transition.

To the last point one might object that the environment should be symmetric and is thus unable to choose a particular classical property. The environment is indeed symmetric but it is only symmetric in an ergodic sense. If we denote by $g$ an element of the symmetry group of the environment and by $\state{\mbox{env}}$ the state of the environment at a particular time then in general we will have
\begin{equation}
g\cdot \state{\mbox{env}} \ne \state{\mbox{env}}.
\end{equation}
The state of the environment is non-symmetric because of the existence of non-symmetric fluctuations. If we define a time averaged state $\state{\mbox{env}, \Delta T}$ by
\begin{equation}
\state{\mbox{env}, \Delta T} = \frac{1}{\Delta T}\int_{\Delta T}dt\ \buni(t) \state{\mbox{env}},
\end{equation}
then we will have indeed 
\begin{equation}
g\cdot \state{\mbox{env}, \Delta T} = \state{\mbox{env}, \Delta T},
\end{equation}
for $\Delta T$ large enough. This is because the non-symmetric fluctuations cancel each other. The presence of these non-symmetric fluctuations together with a transition that is very sensitive to the environment gives rise to the probabilistic appearance of classical properties. 

One can compare this situation to the situation in usual quantum mechanics. In the Copenhagen interpretation of quantum mechanics the environment of the quantum system is provided by a classical observer. It is the contact with this classical observer that leads to the collapse of the wavefunction and the introduction of probability. In our approach the environment is a quantum ensemble. Probability is introduced by the non-symmetric fluctuations of the environment. The obvious advantage of our approach is that no split between observer and system needs to be postulated. The many body system achieves classicality with a quantum environment. 

Probabilistic behavior in the transition from quantum to classical can thus not be avoided. Can one also calculate the corresponding probabilities? Given the structure of Hilbert spaces and inner products that we have assumed from the outset this is indeed possible provided one makes some assumption about the nature of the interaction of the system and the environment. Details can be found in \cite{dreyer}.

\section{Discussion}\label{sec:concl}
In this article we have argued that the measurement problem in quantum mechanics can be traced to the appearance of a classical world in the initial formulation of the theory. If instead a definition of classicality in quantum mechanical terms is adopted the measurement problem can be resolved in the linear setup. The resolution of the problem does not require the introduction of a split between the observer and the object. 

Probabilities arise because the nature of the transition from quantum to classical amplifies the fluctuations in the environment. Probabilities thus have the same origin in quantum mechanics as elsewhere in physics: the lack of knowledge.

The view we are proposing here relies on reasoning common in statistical physics. The world is as it is because this is the most likely way for it to be. The situation in quantum mechanics can be compared to the situation it statistical mechanics at the beginning of the last century. There existed a conceptual tension between the second law of thermodynamics which stated that the entropy of a system never decreases and Poincar\'e recurrence which implied that a system always returns to its initial conditions. No function could be both monotonically increasing and periodic. The solution to this problem was given by the Ehrenfests \cite{ehrenfest} who argued that it was overwhelmingly unlikely for the system to return to its initial state. We have proposed a similar statistical solution to the measurement problem. The occurrence of superpositions of classical objects is not impossible but just very unlikely. John Bell called a solution of this kind a solution for all practical purposes\cite{bell} and introduced the shorthand fapp for it. It is clear that he intended it as a derogatory term. He was after a more fundamental explanation. We have argued that there does not exist a more fundamental solution then the one we have proposed here. We will have to do with a solution fapp.

This view of the foundations of quantum mechanics also sheds light onto the search for a quantum theory of gravity. One lesson is: Do not quantize. If classicality arises as it is proposed here then the more fundamental quantum theory is not obtained by quantizing a classical limit. The search for quantum gravity should thus not start with the classical theory and try to quantize it. Instead one should start with a purely quantum theory. The construction of quantum theories without a classical theory to start with has begun only recently. Examples are the theories constructed by X.-G.~Wen \cite{wen} and S.~Lloyd \cite{lloyd}. A second lesson is that certain concepts that make sense in a classical world might not have a meaning in the more fundamental quantum world. Concepts like locality, position, and momentum might only be strictly applicable and meaningful in a classical context.

\ack 
The author would like to thank Lee Smolin, Chris Beetle, Chris Isham, Fay Dowker, Fotini Markopoulou, Hans Westman, Joy Christian, and Lucien Hardy for comments and discussions. Part of the work presented in this article was carried out at the Perimeter Institute for Theoretical Physics. The author is grateful for the institutes hospitality. This work was presented at the DICE 2006 conference in Piombino. The author thanks Thomas Elze for organizing the workshop and the opportunity to speak.

\end{document}